\documentclass[conference]{IEEEtran}
\ifCLASSINFOpdf
\else

\fi

\usepackage[cmex10]{amsmath}
\usepackage{amssymb}

\usepackage{graphics} 
\usepackage{epsfig}

\usepackage{eso-pic}

\begin{document}
%
\title{Circuit proposition for copying the value of a resistor into a memristive device supported by HSPICE simulation}

\author{\IEEEauthorblockN{Farshad Merrikh-Bayat}
\IEEEauthorblockA{Faculty of Electrical Engineering\\
University of Zanjan\\
Zanjan, Iran, P.O.Box 313\\
Email: f.bayat@znu.ac.ir}\and \IEEEauthorblockN{Nafiseh Mirebrahimi}
\IEEEauthorblockA{Faculty of Electrical Engineering\\
University of Zanjan\\
Zanjan, Iran, P.O.Box 313\\
Email: nafisehmirebrahimi@yahoo.com} \and \IEEEauthorblockN{Farhad Bayat}
\IEEEauthorblockA{Faculty of Electrical Engineering\\
University of Zanjan\\
Zanjan, Iran, P.O.Box 313\\
Email: bayat.farhad@znu.ac.ir}
}

\maketitle

\begin{abstract}
Memristor is the fourth fundamental passive circuit element with
potential applications in development of analog memories,
artificial brains (with the capacity of hardware training) and
neuro-science. In most of these applications the memristance of
the device should be set to the desired value, which is currently
performed by trial and error. The aim of this paper is to propose
a circuit for copying the value of the given resistor into a
memristive device. HSPICE simulations are also presented to
confirm the efficiency of the proposed circuit.
\end{abstract}


%
\IEEEpeerreviewmaketitle

\section{Introduction}
On 1 May 2008 a research group in Hewlett-Packard Labs reported
the physical realization of the first memristor \cite{Strukov},
the fourth fundamental passive circuit element, which was already
predicted by Leon Chua in 1971 \cite{Chua}. After that innovation,
many researchers seek the applications of memristor in variety of
scientific fields such as neuroscience, neural networks and
artificial intelligence. It is now clear that this passive circuit
element can have many applications in, for instance, development
of analog neural networks and emulation of human learning
\cite{ventra2}, building programmable analog circuits
\cite{pershin,Merrikh-bayat}, constructing hardware for soft
computing tools \cite{Merrikh-bayat2}, implementing digital
circuits \cite{Kuekes}, and in the field of signal processing
\cite{Mouttet,Merrikh-Bayat3}.

One main reason for this great interest to memristive computing
systems is that these systems have a high potential to overcome
most of the challenges in front of today's digital systems. For
example, it is proved that these systems can be constructed much
denser than their counterparts through the use of nano-crossbar
technology, and moreover, they consume a considerably less energy
\cite{Snider}.

Roughly speaking, memristor can be considered as a simple resistor
whose resistance is changed by applying voltage or current to it.
This means that an analog value can be stored in this kind of
device by setting its resistance equal to the desired value. It
concludes the fact that memristors can play the role of analog
memories in analog circuits since they can hold their resistance
unchanged until a voltage or a current is applied to them.
However, there is a challenge in front of using memristors as
analog memories. The problem is that we need a simple and
effective circuit for accurate tuning the resistance of the given
memristor to the desired value. This problem has partly been
studied by authors of this paper and the results are presented in
\cite{farshad1}. In that study a circuit for storing an analog
voltage signal in a memristor was proposed. The aim of this paper
is to propose a circuit for automatic adjustment of the resistance
of a memristor to the value of the given resistor.

The rest of this paper is organized as follows. In section
\ref{sec_rev1} we briefly review the notion of memristive systems.
Section \ref{sec_model2} is devoted to the HSPICE model of the
HP-memristor used in our simulations. Two electronic circuits for
automatic tuning the resistance of the HP-memristive to the
desired value are proposed in Section \ref{sec_prop3}. Results of
some HSPICE simulations are are also presented in this Section.
Finally, Section \ref{sec_conc} concludes the paper.

\section{Brief review of memristive systems}\label{sec_rev1}
A \emph{memristive system} can be described by the equations
\begin{equation}\label{mem1}
v(t)=R(\mathbf{x},i)i(t),
\end{equation}
\begin{equation}\label{mem2}
\frac{\mathrm{d}\mathbf{x}}{\mathrm{d}t}=f(\mathbf{x},i),
\end{equation}
where $\mathbf{x}$ is a vector representing internal state
variables of system, $t$ is the time variable, $v(t)$ and $i(t)$
are the voltage and current across the device, respectively, and
$R$ is a scaler called \emph{memristance}
\cite{Strukov,ventra,biolek}. As it can be observed in
(\ref{mem2}), the time derivative of the states of this system
depends on the electrical current passing through it. For this
reason this system is sometimes called the
\emph{current-controlled} memristive system.
\emph{Current-controlled memristor} is a special case of
current-controlled memristive systems. The most basic mathematical
definition of a current-controlled memristor is the differential
form \cite{Strukov}:
\begin{equation}
v=R(x)i,
\end{equation}
\begin{equation}
\frac{\mathrm{d}x}{\mathrm{d}t}=i.
\end{equation}
Hence, according to (\ref{mem1}) and (\ref{mem2}) the
current-controlled memristor can be considered as a kind of
(time-varying) resistor whose resistance depends on the history of
the current passed through it and also to the internal state of
device.

\begin{figure}[tb]
\begin{center}
\includegraphics[width=1.5cm]{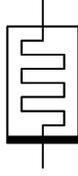}
\caption{The symbol of memristor} \label{fig_mem}
\end{center}
\end{figure}

Figure \ref{fig_mem} shows the symbol of memristor used in the
literature. The memristor shown in this figure is an asymmetric
device which has the property that applying positive voltage to
the terminal denoted by the black thick line with respect to other
terminal decreases its resistance, and vice versa. It concludes
that in practice the memristance of a memristor can be adjusted to
the desired value by applying a suitable alternating-polarity
voltage to it \cite{Alibart} (note that in practice the
memristance of memristive devices always lies between two limiting
values denoted as $R_{\mathrm{on}}$ and $R_\mathrm{off}$, where
$R_\mathrm{on}<R_\mathrm{off}$ and the ratio of these two
resistances is usually given as $10^2$-$10^3$). Since such an
appropriate voltage generator is not available at this time, in
practice the memristance of the given memristor is adjusted to the
desired value by trial and error. More precisely, positive and
negative voltages are applied to the memristor to decrease and
increase its memristance respectively until it takes a value close
to the desired one. The main drawback of this approach is that it
is commonly time consuming, and moreover, the results may not be
accurate enough.

In the following, we will propose a circuit for automatic and
accurate tuning the memristance of the given memristor to the
desired value in a reasonably short time. Since we will use the
model of the memristor constructed in HP labs in our simulations,
first we need to present the HSPICE model of this circuit element.

\section{HSPICE model of HP-memristor}\label{sec_model2}
The HSPICE model of the memristor used in simulations of this
paper is based on the physical model proposed for memristor in
\cite{Strukov} (which is currently known as the HP-memristor) and
the corresponding electrical model proposed for it in
\cite{biolek}. In the physical model proposed in \cite{Strukov},
the memristor is considered as a two-layer thin film of
$\mathrm{TiO_2}$ as shown in Fig. \ref{fig_mem2} ($D\approx 10$
nm), which is sandwiched between two platinum contacts. One of the
layers in this figure is doped with oxygen vacancies (and
consequently, behaves as a semiconductor) and the other is undoped
(and consequently, behaves an insulator). The boundary between two
layers is moved in the same direction as the electrical current
passes. It is concluded from Fig. \ref{fig_mem2} that the total
resistance of the memristor, $R_\mathrm{mem}(x)$, is equal to the
sum of the resistances of doped and undoped regions as follows:
\begin{equation}
R_\mathrm{mem}(x)=R_\mathrm{on}x+R_\mathrm{off}(1-x)=R_\mathrm{off}-(R_\mathrm{off}-R_\mathrm{on})x,
\end{equation}
where
\begin{equation}
x=\frac{w}{D}\in(0,1),
\end{equation}
and $R_\mathrm{off}$ and $R_\mathrm{on}$ (which correspond to
$w=0$ and $w=D$, respectively) are equal to the maximum and
minimum possible values for the resistance of memristor,
respectively. In fact, as mentioned before, the resistance of the
memristor is always between the limiting values $R_\mathrm{on}$
and $R_\mathrm{off}$. According to the discussions presented in
\cite{Strukov} and \cite{biolek} the derivative of $x$ with
respect to time can be considered as the following:
\begin{equation}\label{dx}
\frac{\mathrm{d}x}{\mathrm{d}t}=ki(t)f(x), \quad k=\frac{\mu_v
R_{on}}{D^2},
\end{equation}
where $\mu_v\approx 10^{-14}\mathrm{m^2s^{-1}V^{-1}}$ and $f(x)$
is the so-called \emph{window function}. The window function
$f(x)$ in (\ref{dx}) can be defined in many different ways. One
possible approach is to define it as \cite{biolek,joglekar}:
\begin{equation}
f(x)=1-(2x-1)^p,
\end{equation}
where $p$ is a positive integer.

\begin{figure}[tb]
\begin{center}
\includegraphics[width=4.5cm]{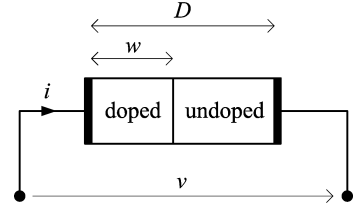}
\caption{The physical model of memristor} \label{fig_mem2}
\end{center}
\end{figure}

Based on the above discussions and the approach proposed in
\cite{biolek} for electrical modeling of memristive devices, the
HSPICE model presented in Table 1 is considered to be used in
simulations of this paper (this model is the same as the one
proposed in \cite{biolek} with small modifications to make it
compatible with HSPICE simulator). In this table, which
approximately models the HP-memristor, it is assumed that
$R_\mathrm{on}=100\Omega$, $R_\mathrm{off}=16\mathrm{k}\Omega$,
$p=10$, and the memristance
of the device is initially equal to $1\mathrm{k}\Omega$.\\

Table 1. HSPICE model of the HP-memristor used in simulations of
this paper (the fifth line of this code is the continuation of
fourth line and during the simulation these two lines should be
typed in a same line).\\
\begin{tabular}{|l|}
  \hline
  ******** HP memristor model ********\\
    .SUBCKT memristor Plus Minus\\
    + Ron=100 Roff=16K Rinit=1K D=10N uv=10F p=10\\
    Gx 0 x
    CUR='(I(Emem)*(uv*Ron))/(pow(D,2))*\\(1-pow((2*V(x)-1),(2*p)))'\\
    Cx x 0 1 IC='(Roff-Rinit)/(Roff-Ron)'\\
    Raux x 0 1T\\
    Emem plus aux VOL='-I(Emem)*V(x)*(Roff-Ron)'\\
    Roff aux minus 'Roff'\\
    .ENDS\\
  \hline
\end{tabular}\label{hspice}

\section{Proposed circuit for copying the resistance of the given resistor in a memristor and HSPICE
simulation}\label{sec_prop3} 
\begin{figure}[tb]
\begin{center}
\includegraphics[width=7.5cm]{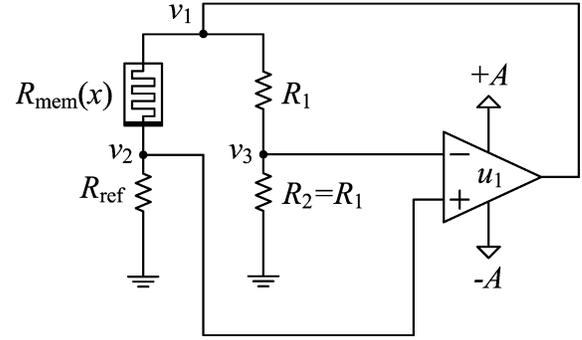}
\caption{The proposed circuit for increasing the resistance of
memristor from $R_{\mathrm{mem}}$ to $R_{\mathrm{ref}}$.}
\label{fig_cir1}
\end{center}
\end{figure}

Figure \ref{fig_cir1} shows the
proposed circuit for increasing the resistance of the given
memristor from $R_{\mathrm{mem}}$ to the desired value
$R_{\mathrm{ref}}$ ($R_{\mathrm{ref}}
>R_{\mathrm{mem}}$). More precisely, in this figure it is assumed
that the initial value of the memristor is smaller than
$R_{\mathrm{ref}}$ and the circuit has the property that increases
the resistance of the memristor to the desired value
$R_{\mathrm{ref}}$. In Fig. \ref{fig_cir1} the value of $R_1$ must
be equal to $R_2$ and theoretically, they can be considered equal
to any number. Any Op-Amp can also be used in this connection. All
simulations of this paper are performed assuming
$R_1=R_2=1\mathrm{k}\Omega$, when the AD711a Op-Amp is applied and
$A=5$V.

Analysis of the circuit shown in Fig. \ref{fig_cir1} is
straightforward. Considering the fact that $R_1=R_2$, we always
have $v_3=v_1/2$. On the other hand, since at the beginning we
have $R_{\mathrm{ref}} >R_{\mathrm{mem}}$, and consequently
$v_2>v_1/2$, the output of Op-Amp is at the high state, which
leads to increasing the resistance of memristor. By increasing the
resistance of memristor, $v_2$ is also increased which results in
decreasing the output voltage of Op-Amp. At steady state, we have
$v_2=v_3$, $v_1=0$, and $R_\mathrm{mem}(x)=R_\mathrm{ref}$.

\begin{figure}
\begin{center}
\includegraphics[width=8.5cm]{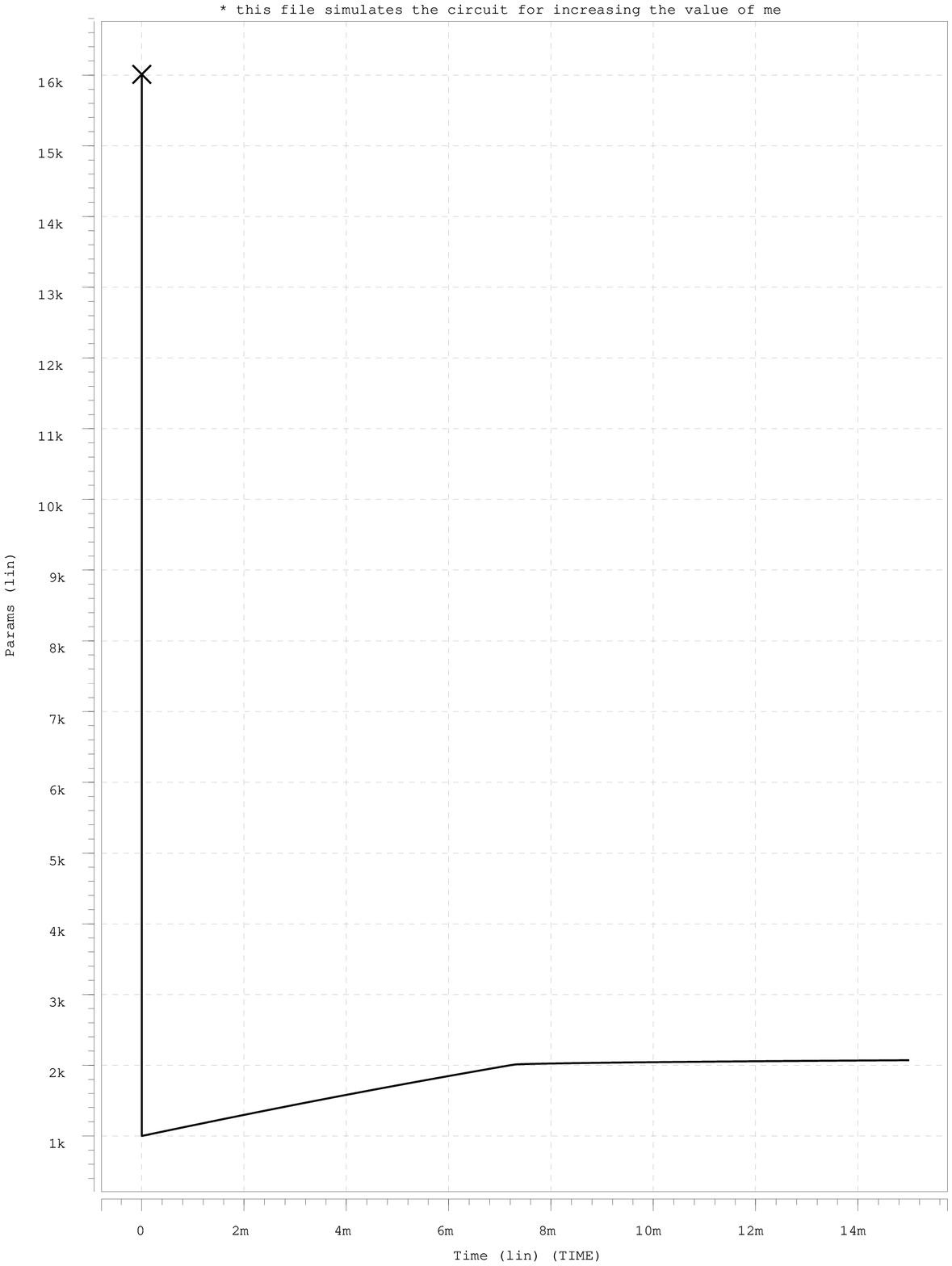}
\caption{Resistance of the memristor of Fig. \ref{fig_cir1} versus
time when $R_{\mathrm{ref}}=2\mathrm{k}\Omega$ and the initial
value of memristor is equal to 1k$\Omega$. Memristance of the
device tends to $R_{\mathrm{ref}}$ by increasing the time.}
\label{fig_mem1}
\end{center}
\end{figure}

\begin{figure}
\begin{center}
\includegraphics[width=8.5cm]{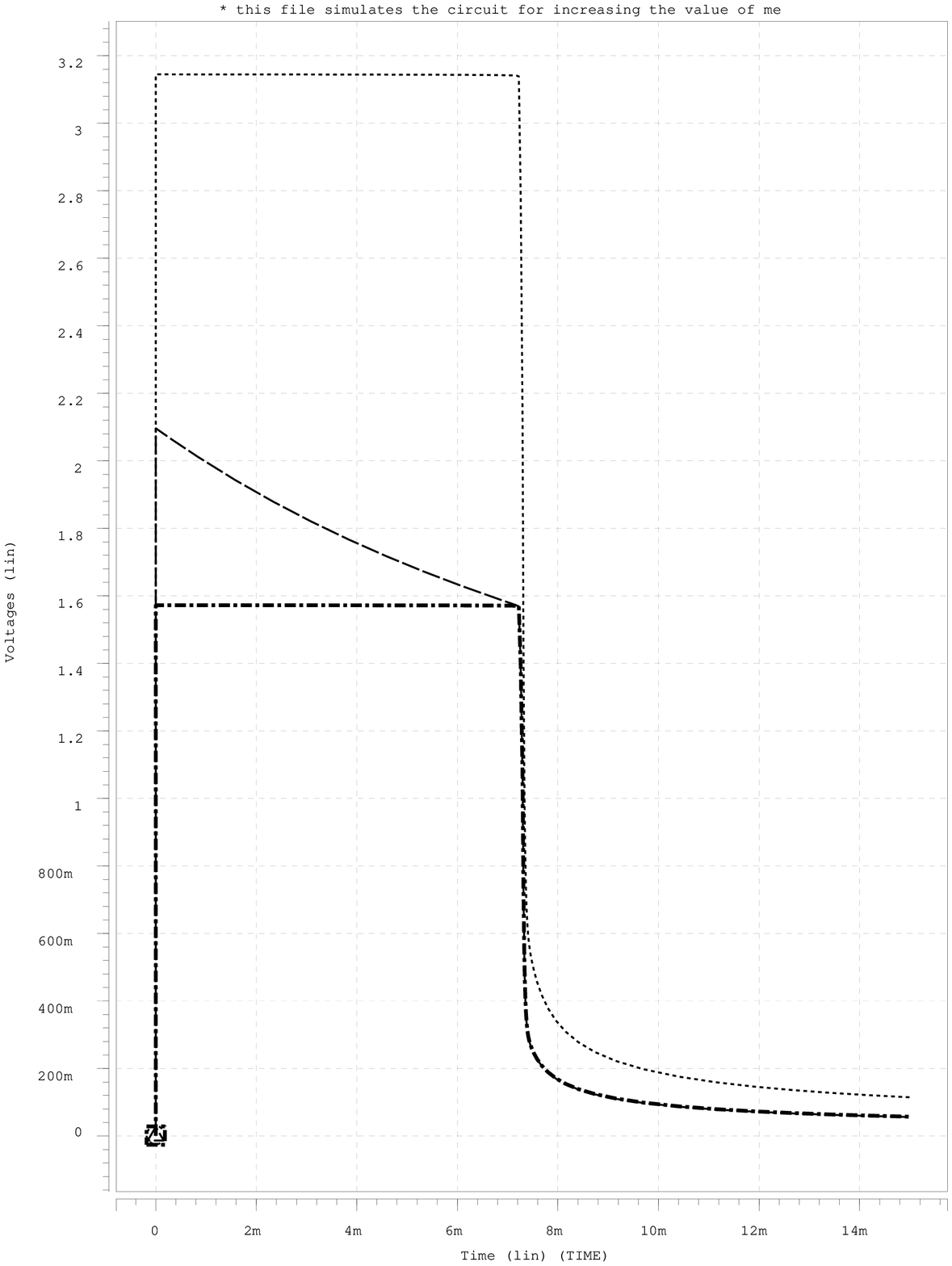}
\caption{The plot of $v_1$ (dotted curve), $v_2$ (dashed curve),
and $v_3$ (dashed-dotted curve) in Fig. \ref{fig_cir1} versus time
($R_1=R_2=1\mathrm{k}\Omega$,
$R_{\mathrm{ref}}=2\mathrm{k}\Omega$, and $A=5$V).}
\label{fig_vol1}
\end{center}
\end{figure}

\begin{figure}
\begin{center}
\includegraphics[width=8.5cm]{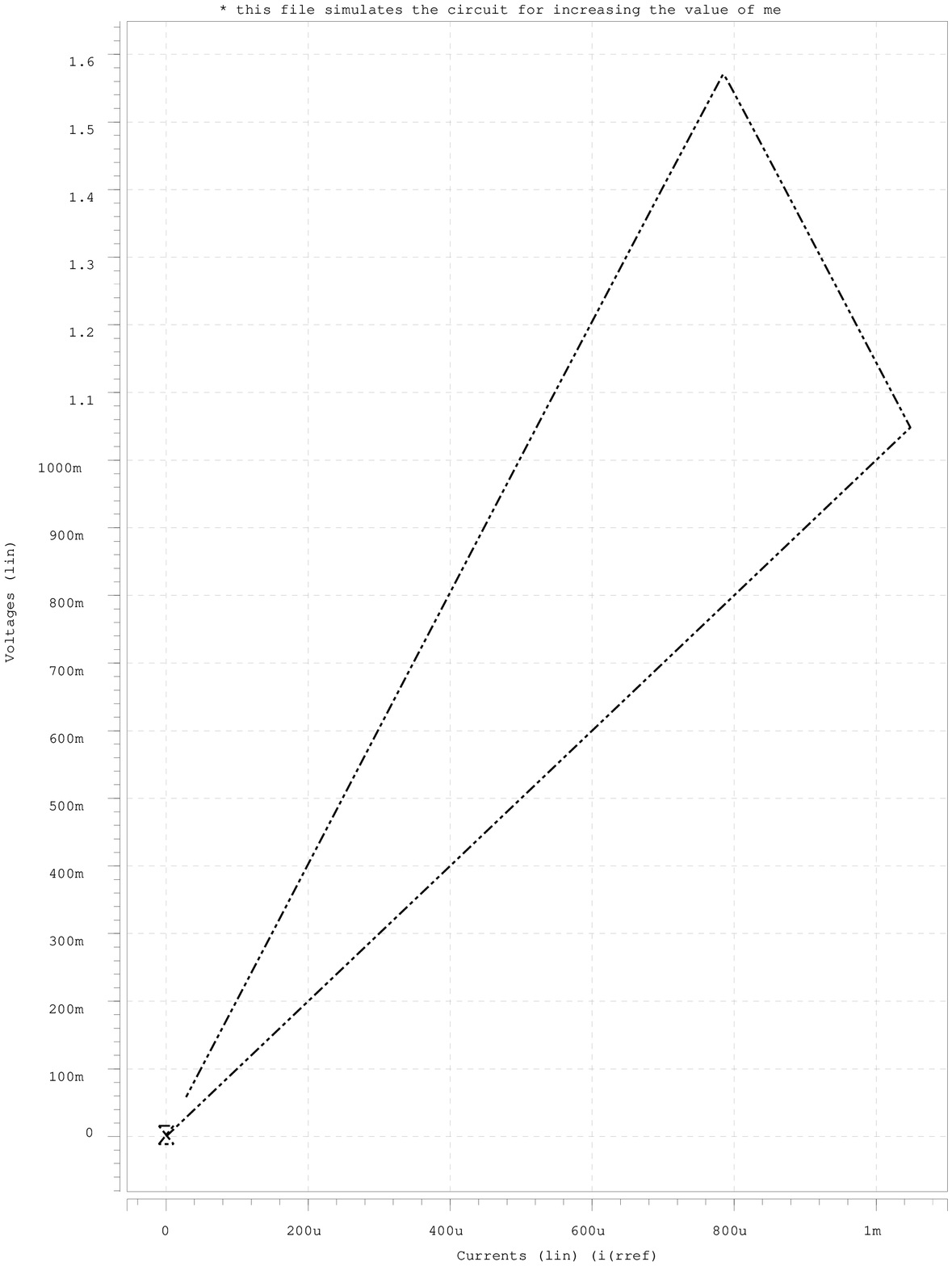}
\caption{Plot of the voltage across memristor versus the current
pass through it in Fig. \ref{fig_cir1}
($R_1=R_2=1\mathrm{k}\Omega$,
$R_{\mathrm{ref}}=2\mathrm{k}\Omega$, and $A=5$V). The slope of
plot at any point indicates the memristance of device at that
point.} \label{fig_ph1}
\end{center}
\end{figure}

Figures \ref{fig_mem1}-\ref{fig_ph1} show the HSPICE simulations
of the circuit shown in Fig. \ref{fig_cir1}. All of these
simulations are done in the time range 0-15ms, where the time step
is considered equal to $1$ns. Note that larger time steps may
result in inaccurate results or even divergence of the solutions.
Another point that should be noted during simulation is that the
port of memristor with plus sign in Table 1 corresponds to the
port with black thick line in Fig. \ref{fig_cir1}. Figure
\ref{fig_mem1} shows the resistance of the memristor of Fig.
\ref{fig_cir1} versus time. As it can be observed, the memristance
of the device tends to $R_{\mathrm{ref}}=2\mathrm{k}\Omega$ as the
time is increased. A small steady-state error observed in Fig.
\ref{fig_mem1} may because of the value considered for time step
since this error is decreased by decreasing the time step used for
simulation. In this simulation the memristance reaches its
steady-state value after about 7ms. Simulations show that this
time is increased by increasing $R_{\mathrm{ref}}$ or decreasing
the power supply applied to Op-Amp (i.e., the value of $A$ in Fig.
\ref{fig_cir1}). For example, Fig. \ref{fig_mem2} shows the
resistance of the memristor of Fig. \ref{fig_cir1} versus time
when $R_{\mathrm{ref}}=4\mathrm{k}\Omega$. This figure clearly
shows that increasing $R_{\mathrm{ref}}$ from 2k$\Omega$ to
4k$\Omega$ increases the settling time of the system response from
7ns to 40ns.

\begin{figure}
\begin{center}
\includegraphics[width=8.5cm]{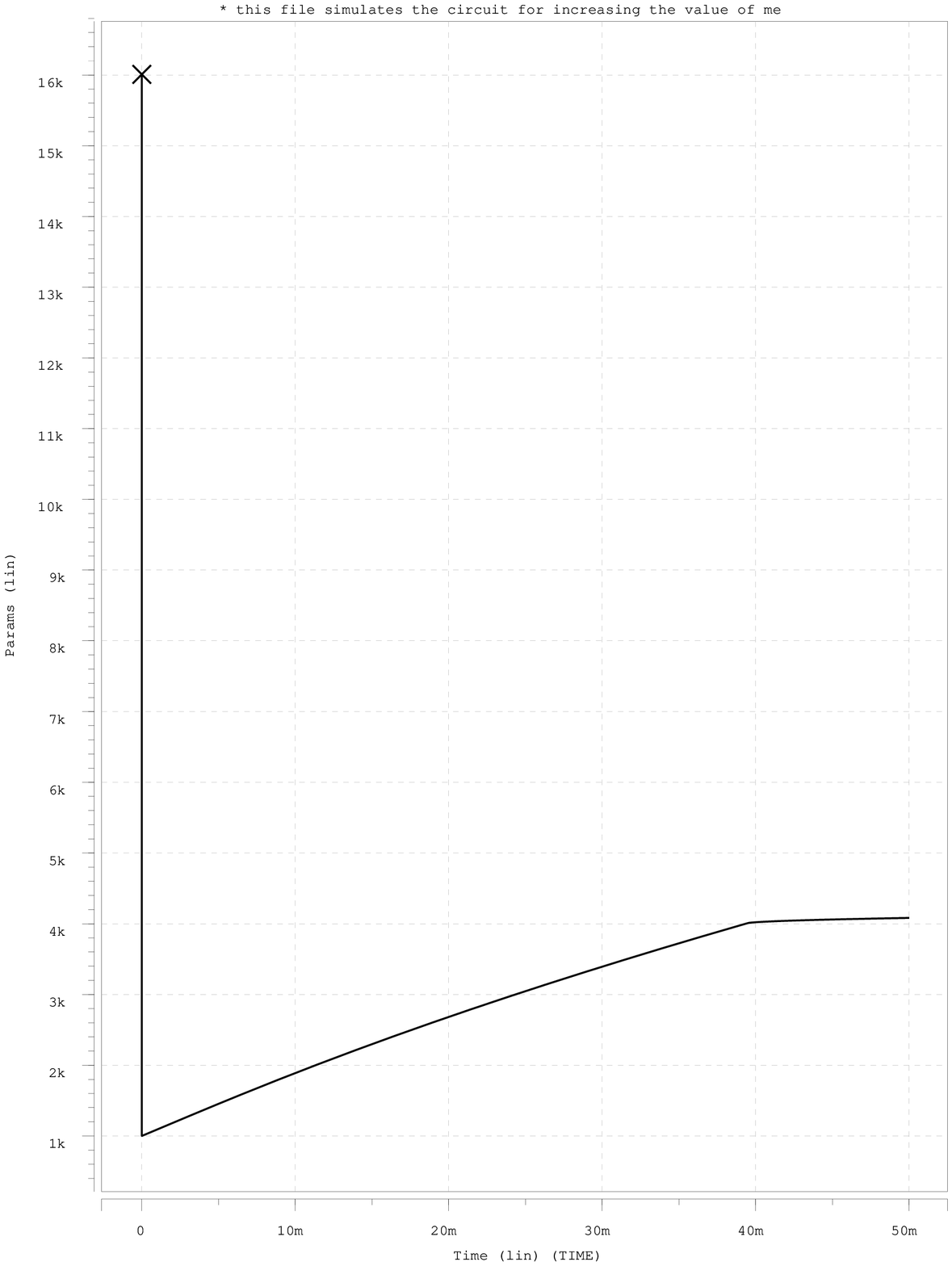}
\caption{Resistance of the memristor of Fig. \ref{fig_cir1} versus
time when $R_{\mathrm{ref}}=4\mathrm{k}\Omega$ and the initial
value of memristor is equal to $1\mathrm{k}\Omega$.}
\label{fig_mem2}
\end{center}
\end{figure}

\begin{figure}[tb]
\begin{center}
\includegraphics[width=7.5cm]{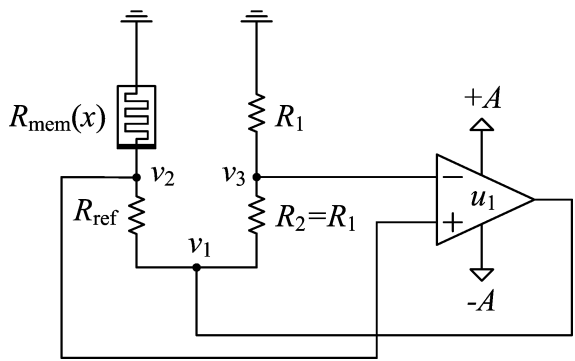}
\caption{The proposed circuit for decreasing the resistance of
memristor from $R_{\mathrm{mem}}$ to $R_{\mathrm{ref}}$}
\label{fig_cir2}
\end{center}
\end{figure}

It can be easily verified that the circuit shown in Fig.
\ref{fig_cir1} does not work when $R_{\mathrm{ref}}
<R_{\mathrm{mem}}$, that is, this circuit cannot be used for
decreasing the memristance of the given memristive device. In such
cases the circuit shown in Fig. \ref{fig_cir2} can be used
instead. The function of this circuit can be explained similar to
the circuit shown in Fig. \ref{fig_cir1} (note to the similarities
between these two circuits).

\begin{figure}[tb]
\begin{center}
\includegraphics[width=8.5cm]{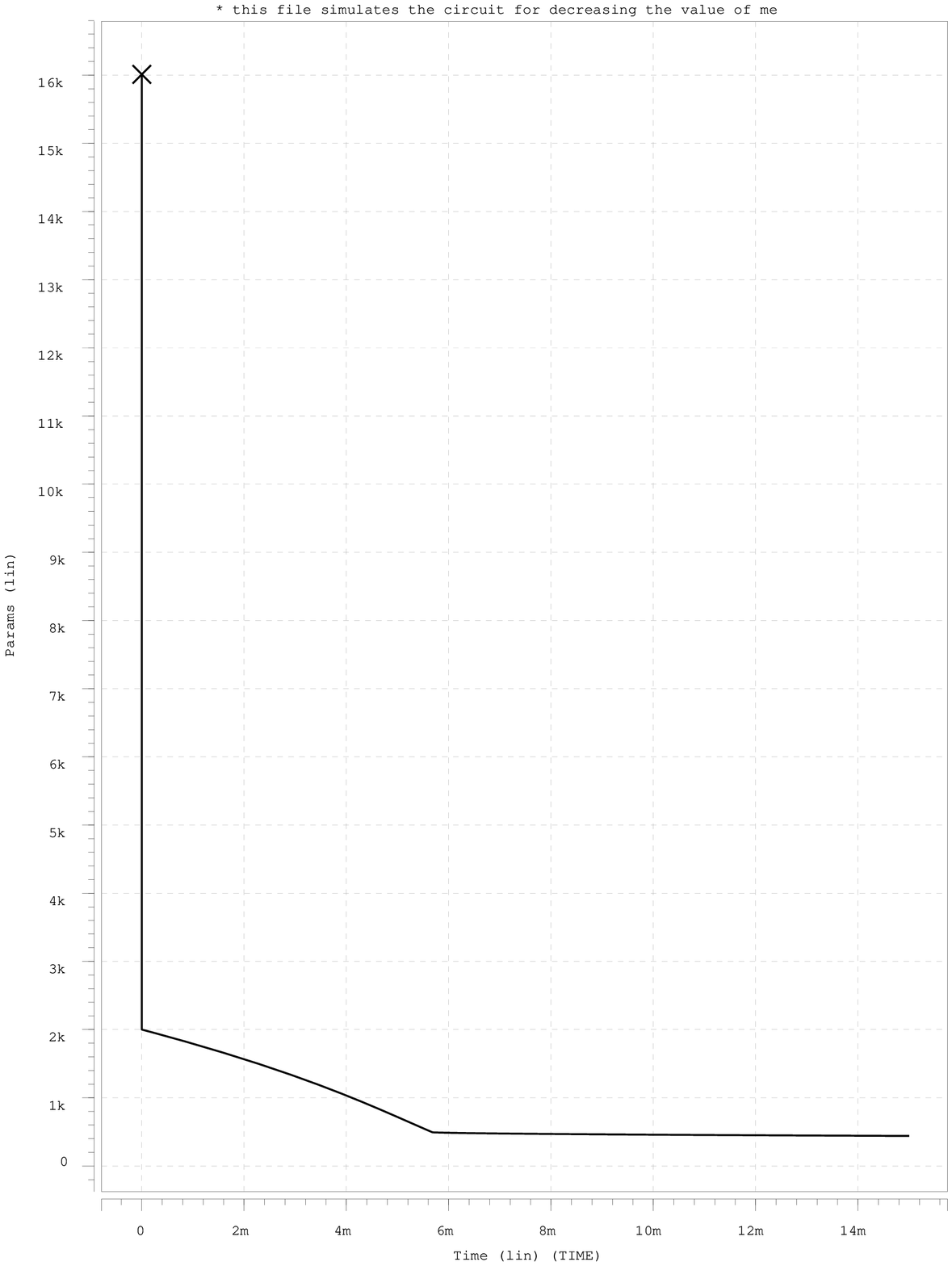}
\caption{Resistance of the memristor of Fig. \ref{fig_cir2} versus
time when $R_{\mathrm{ref}}=500\Omega$ and the initial value of
memristor is equal to $2\mathrm{k}\Omega$. Memristance of the
device tends to $R_{\mathrm{ref}}$ by increasing the time.}
\label{fig_mem3}
\end{center}
\end{figure}

\begin{figure}[tb]
\begin{center}
\includegraphics[width=8.5cm]{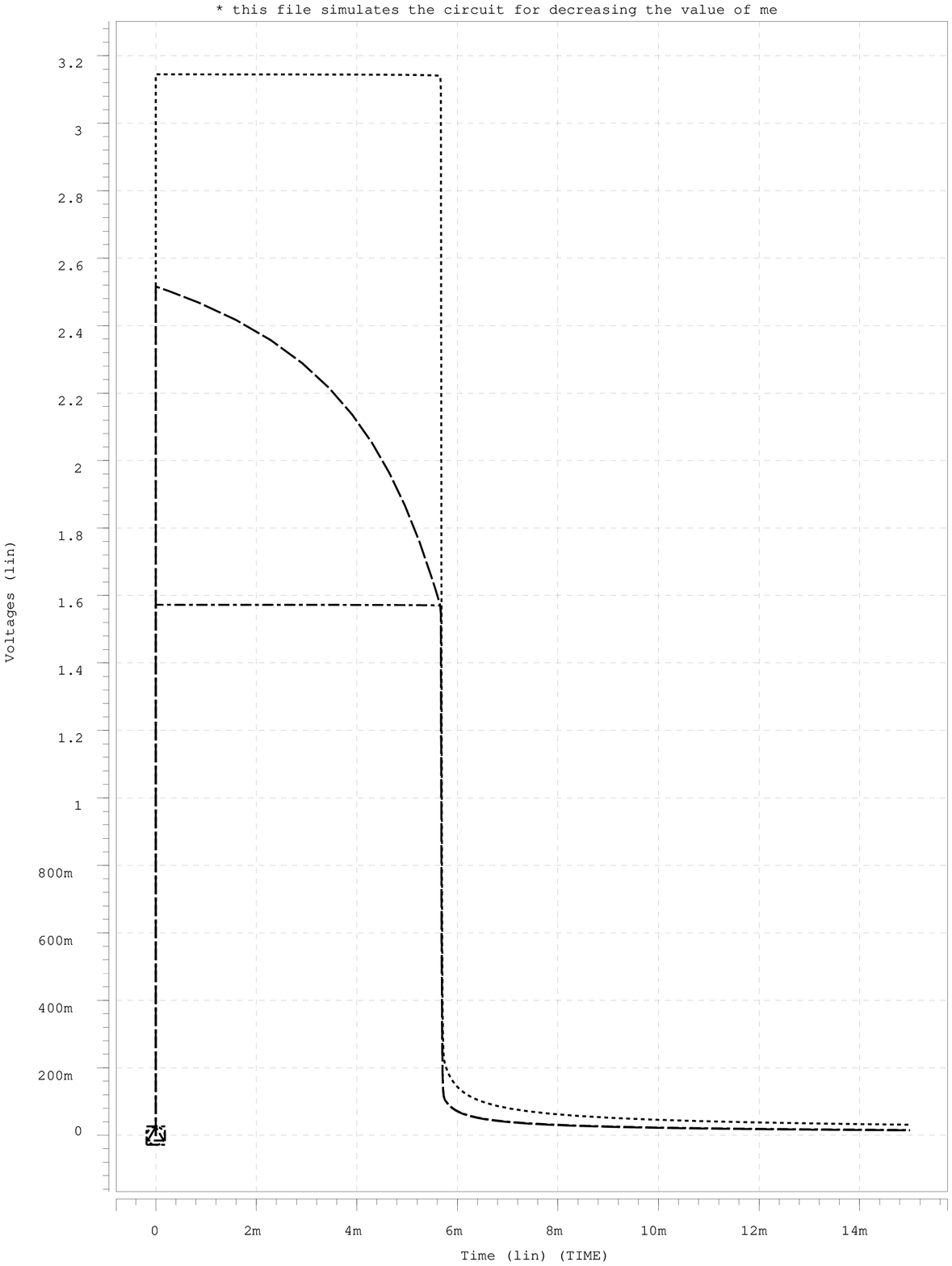}
\caption{The plot of $v_1$ (dotted curve), $v_2$ (dashed curve),
and $v_3$ (dashed-dotted curve) in Fig. \ref{fig_cir2} versus time
($R_1=R_2=1\mathrm{k}\Omega$, $R_{\mathrm{ref}}=500\Omega$, and
$A=5$V).} \label{fig_vol2}
\end{center}
\end{figure}

\begin{figure}[tb]
\begin{center}
\includegraphics[width=8.5cm]{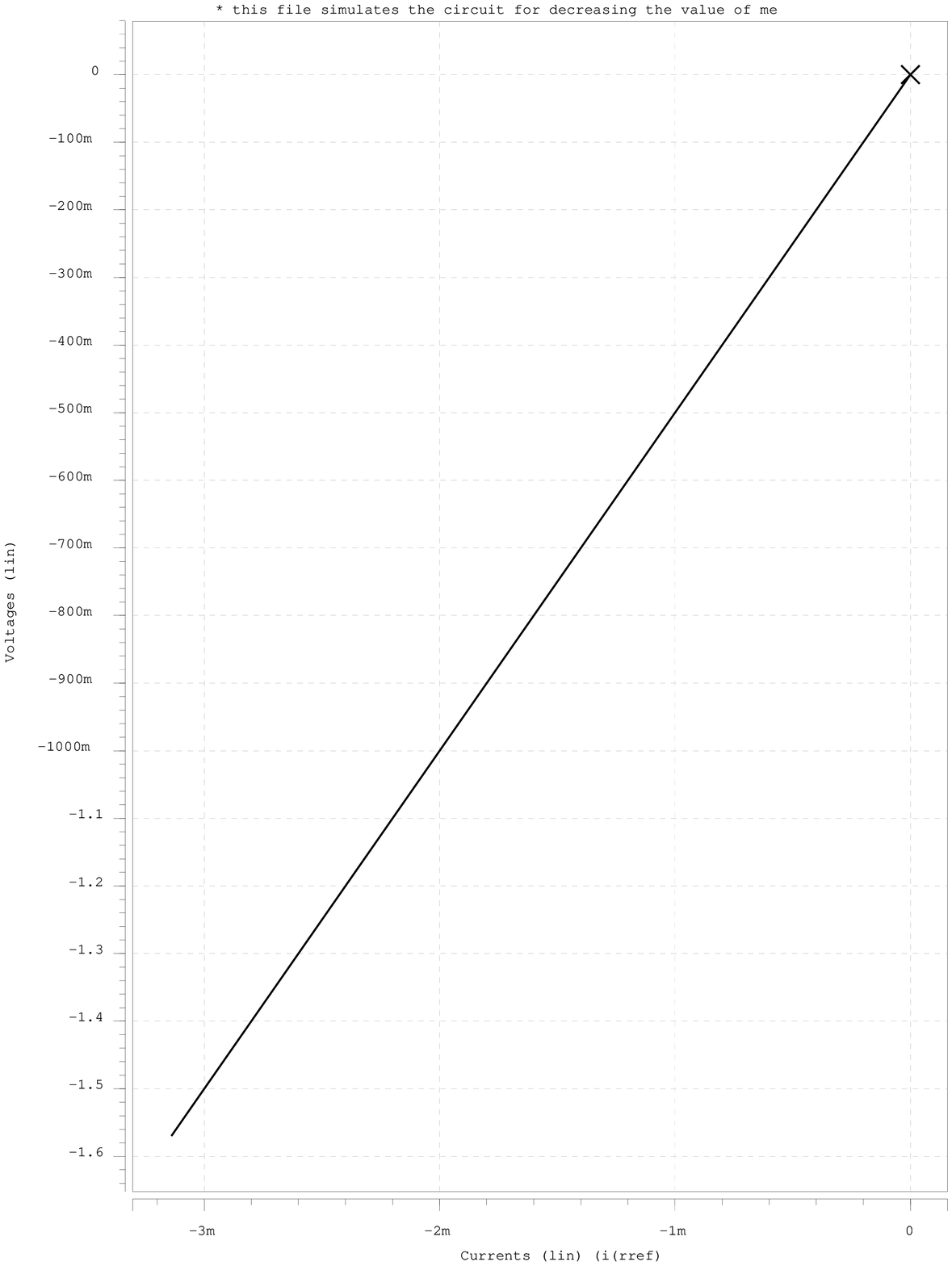}
\caption{Plot of the voltage across memristor versus the current
pass through it in Fig. \ref{fig_cir2}
($R_1=R_2=1\mathrm{k}\Omega$, $R_{\mathrm{ref}}=500\Omega$, and
$A=5$V). The slope of plot at all points is approximately equal to
0.5, which corresponds to the resistance of $500\Omega$ for
memristor, as it is expected.}\label{fig_ph2}
\end{center}
\end{figure}

Figures \ref{fig_mem3}-\ref{fig_ph2} show the simulation results
of the circuit shown in Fig. \ref{fig_cir2}. Similar to the
previous simulations these results are also obtained assuming
$R_1=R_2=1\mathrm{k}\Omega$, $A=5$V, and the time step of 1ns,
while it is assumed that $R_{\mathrm{ref}}=500\Omega$ and the
resistance of memristor is initially equal to $2\mathrm{k}\Omega$.
Figure \ref{fig_mem3} clearly shows that the proposed connection
can effectively decrease the memristance of the device until it
becomes equal to the desired value $R_{\mathrm{ref}}$. Figure
\ref{fig_vol2} shows the voltage signals corresponding to the
simulation of Fig. \ref{fig_mem3}. As it can be observed in this
figure, $v_2$ and $v_3$ become equal after less than 6ms. Figure
\ref{fig_ph2} shows the plot of voltage across memristor in Fig.
\ref{fig_cir2} versus the corresponding current pass through it.
The slope of this curve at any point on it indicates the
resistance of the memristor at that point. It can be easily
concluded from this figure that the memristance of device tends to
$500\Omega$ as the time is increased. Note that the differences
between the simulation results presented in Figs. \ref{fig_ph1}
and $\ref{fig_ph2}$ are mainly because of two reasons: First, the
memristor is an asymmetric device, and second, these two
simulations are performed assuming different initial and final
values.

\section{Discussion and Conclusion}\label{sec_conc}
Two circuits for adjusting the resistance of a memristor to the
value of the given resistor are proposed in this paper, and HSPICE
simulations are presented. One of the proposed circuits can be
used for increasing the memristance of the device to the desired
value and the other one can be used for decreasing it. These
circuits can be used for adjustment of the resistance of the given
memristor to the desired value and copying the value of a resistor
in the memristor.

There are, however, many other questions that can be considered as
the subject of future studies. Some of them are listed below:
\begin{itemize}
    \item The proposed circuits can be implemented by using few
    transistors instead of applying an Op-Amp. In fact, there is
    no need to use an Op-Amp, which may consist of tens of
    transistors, in these circuits. More simple designs are
    desired and can be considered as the subject of future work.
    \item We proposed two circuits in this paper: one for
    increasing the memristance of the device and the other for
    decreasing it. In practice it is highly demanded to have a
    unique circuit with the ability of automatic adjustment of the memristance of the device
    to the desired value either by increasing or decreasing its
    resistance. Development of such a circuit (possibly by combining
    the proposed two circuits) can be considered as the subject of
    future studies.
    \item Intuitively, it is expected that the proposed circuits can be
    used for automatic adjustment of the resistance of any
    memristive device (in addition to the HP-memristor studied in this paper). It is
    important to provide a mathematical proof for this statement
    and exactly determine the possible limitations of the proposed designs.
    It may even be observed that the application of proposed
    circuits is limited to a certain class of memristive devices.
    Since the circuits studied in this paper are nonlinear in nature, the Lyapunov
    stability theorem \cite{slotine} can be used for this purpose.
    \item All simulations of this paper are carried out assuming that the reference
    resistor is constant (i.e., it is not varied with time). Although it may also be
    expected that the proposed circuits can handle time-varying
    reference resistors, it needs mathematical proof and numerical
    simulation, which is not performed in this paper.

    \item Another very important issue is to design a simple
    circuit for copying the analog value stored in a memristive device
    (in the form of its resistance) into another memristor. Such
    a circuit is useful in dealing with analog memories
    implemented on crossbar structures.
\end{itemize}


%
%

%
%



\begin{thebibliography}{1}


\bibitem{Strukov}
D.B. Strukov, G.S. Snider, D.R. Stewart, and R.S. Williams, The
Missing Memristor Found, {\it Nature}, vol. 453, pp. 80--83, 1 May
2008.

\bibitem{Chua}
L.O. Chua, Memristor - The Missing Circuit Element, {\it
IEEE Trans. on Circuit Theory}, vol. CT-18, no. 5, pp. 507--519, 1971.

\bibitem{ventra2}
Y.V. Pershin, S. La Fontaine, and M. Di Ventra, Memristive Model
of Amoeba's Learning, \emph{Phys. Rev. E}, vol. 80, pp.
021926/1--021926/6, 2009.

\bibitem{pershin}
Y.V. Pershin and M.D. Ventra, Practical Approach to Programmable
Analog Circuits with Memristors, \emph{IEEE Transactions on
Circuits and Systems I: Regular Paper}, vol. 57, no. 8, pp.
1857--1864, 2010.

\bibitem{Merrikh-bayat}
F. Merrikh-Bayat and S.B. Shouraki, Memristor-Based Circuits for
Performing Basic Arithmetic Operations, \emph{Procedia-Computer
Science Journal}, vol. 3,  pp. 128--132, 2011.


\bibitem{Merrikh-bayat2}
F. Merrikh-Bayat and S.B. Shouraki, Memristor Crossbar-Based
Hardware Implementation of IDS Method, \emph{IEEE Transaction on
Fuzzy Systems}, vol. 19, no. 6,  pp. 1083--1096, 2011.


\bibitem{Kuekes}
P. Kuekes, Material Implication: Digital Logic with Memristors, in
\emph{Proceedings of the Memristor and Memristive Systems
Symposium}, 21 November 2008.

\bibitem{Mouttet}
B.L. Mouttet, Proposal for Memristors in Signal Processing, in
\emph{Proceedings of the Nano-Net Conference}, vol. 3, pp. 11--13,
Sept. 2008.

\bibitem{Merrikh-Bayat3}
F. Merrikh-Bayat and S.B. Shouraki, Mixed Analog-Digital
Crossbar-Based Hardware Implementation of Sign-Sign LMS Adaptive
Filter, \emph{Analog Integrated Circuits and Signal Processing},
vol. 3, no. 1, pp. 41--48, 2011.

\bibitem{Snider}
G. Snider, R. Amerson, D. Carter, H. Abdalla, M.S. Qureshi, J.
Leveille, M. Versace, H. Ames, S. Patrick, B. Chandler, A.
Gorchetchnikov, and E. Mingolla, From Synapses to Circuitry: Using
Memristive Memory to Explore the Electronic Brain, \emph{IEEE
Computer}, vol. 44, no. 2, pp. 21--28, February 2011.

\bibitem{farshad1}
F. Merrikh-Bayat, F. Merrikh-Bayat, and N. Mirebrahimi, A Method
for Automatic Tuning the Memristance of Memristive Devices with
the Capacity of Applying to Memristive Memories, in
\emph{Proceedings of the IEEE International Conference on Computer
Systems and Industrial Informatics}, Dec. 18-20, 2012, Sharjah,
UAE.

\bibitem{ventra}
M. Di Ventra, Y.V. Pershin, and L.O. Chua, Circuit Elements with
Memory: Memristors, Memcapacitors, and Meminductors, {\it
Proceedings of the IEEE}, vol. 97, no. 10,  pp. 1717-–1724, 2009.

\bibitem{biolek}
D. Biolek, Z. Biolek, and V. Biolkova, SPICE Modeling of
Memristive, Memcapacitative and Meminductive Systems, in
\emph{Proceedings of the European Conference on Circuit Theory and
Design - ECCTD}, pp. 249-252, 2009.

\bibitem{Alibart}
F. Alibart, L. Gao, B. Hoskins, and D.B. Strukov, High-Precision
Tuning of State for Memristive Devices by Adaptable
Variation-Tolerant Algorithm, {\it Nanotechnology}, vol. 23, art.
075201, 2012.

\bibitem{joglekar}
Y.N. Joglekar and S.J. Wolf, The Elusive Memristor: Properties of
Basic Electrical Circuits, arXiv: 0807.3994 v2 [cond-mat.mes-hall]
13 January 2009, pp. 1-24.


\bibitem{slotine}
J.E. Slotine and W. Li, \emph{Applied Nonlinear Control}, Prentice
Hall, Englewood Cliffs, N.J., 1991.







%
%







\end{thebibliography}
%

\end{document}